\documentclass{article}
\usepackage{epsfig,amssymb}

\def\beq{\begin{equation}}
\def\eeq{\end{equation}}
\def\bea{\begin{eqnarray}}
\def\eea{\end{eqnarray}}
\def\d{{\mathrm{d}}}

\newfont{\cursive}{pzcmi at 9pt}
\def\~t{\tilde{t}}

\title{The spatial relation between the event horizon and trapping horizon.}
\author{Alex B. Nielsen \\ Max-Planck-Institut f\"ur
    Gravitationsphysik, \\ Albert-Einstein-Institut, \\ Am M\"uhlenberg 1,
    D-14476 Golm, \\ Germany}

\begin{document}

\maketitle

\begin{abstract}
The relation between event horizons and trapping horizons is investigated in a number of different situations with emphasis on their role in thermodynamics. A notion of constant change is introduced that in certain situations allows the location of the event horizon to be found locally. When the black hole is accreting matter the difference in area between the two different horizons can be many orders of magnitude larger than the Planck area. When the black hole is evaporating the difference is small on the Planck scale. A model is introduced that shows how trapping horizons can be expected to appear outside the event horizon before the black hole starts to evaporate. Finally a modified definition is introduced to invariantly define the location of the trapping horizon under a conformal transformation. In this case the trapping horizon is not always a marginally outer trapped surface.

\end{abstract}


\section{Introduction}

Black holes are defined by their horizons. There are different types of black hole horizons that can be used in different situations. What type of horizon is relevant may depend on the question being asked. In black hole thermodynamics the area of the black hole plays a role analogous to entropy via the Bekenstein-Hawking relation. There has been a rigorous attempt to understand whether this entropy has an underlying microscopic explanation and it is hoped that the answer to this question will be an important clue to a theory of quantum gravity. 

The event horizon of future null infinity is most often associated with the boundary of a black hole. It has however been suggested that trapping horizons play an important role in black hole thermodynamics. It is possible that it is the trapping horizon area and not the area of the event horizon that should be considered in black hole thermodynamics \cite{Hiscock:1989uj,Collins:1992uj,Nielsen:2008cr} (for a dissenting view see \cite{Sorkin:1996sr,Corichi:2000xf}). This raises a question of how different the areas of the event horizon and trapping horizon can be. If the difference in areas is sufficiently large, much larger than the fundamental area in Planck units, then a theory of quantum gravity should be able to tell us which area contains the black hole entropy.

Furthermore in theories of semi-classical gravity with quantum fields on a classical spacetime background, one expects black holes to emit Hawking radiation. It has also been suggested that the trapping horizon plays a role in generating Hawking radiation \cite{Di Criscienzo:2007fm,Nielsen:2008kd} (for a dissenting view see \cite{Barcelo:2006uw}). In general numerical simulations the trapping horizon always lies inside the event horizon. If the radiation comes directly from the trapping horizon and the trapping horizon is hidden by the event horizon, then the radiation will not be visible from infinity. In fact, no physical effects of the trapping horizon will be visible from outside.

A theorem of Hawking and Ellis \cite{Hawking:1973uf} implies that the apparent horizon always lies behind the event horizon. In such cases the apparent horizon is never visible to asymptotic observers and its physical properties cannot influence them. The theorem depends on the validity of the null energy condition in the future of the region of the event horizon. It is the same null energy condition that implies that the area of the event horizon cannot decrease. If the area of the event horizon is to decrease, a scenario that is envisioned in black hole evaporation through Hawking radiation, then null energy condition must be violated somewhere along the event horizon. The quantum fields responsible for the Hawking radiation must be capable of violating the null energy condition. In fact this scenario is borne out in explicit calculations in certain simplified models \cite{Visser:1996iw}.

If Hawking radiation violates the null energy condition then in certain stages of its lifetime the apparent horizon can appear outside the event horizon. However, the null energy condition does not need to be violated locally in order for this to happen. It is sufficient that it be violated somewhere in the future, reflecting the teleological nature of the event horizon. In this paper we give an explicit model to show this effect.

Although quantum effects are implied by both these motivations, the relation between the event horizon and trapping horizons can be studied in purely classical relativity on Lorentzian manifolds. In principle the difference in areas can be arbitrarily large. For example there are spacetimes that contain trapping horizons but not event horizons \cite{Roman:1983zza,Hayward:2005gi}. In some other models \cite{Ashtekar:2005qt} there are no event horizons but there is a region in which causally propagated signals cannot avoid a quantum non-manifold region. In such cases a ``quantum horizon'' may replace the notion of an event horizon. In the standard semi-classical model of black hole formation and evaporation \cite{Hawking:1974sw} there is both an apparent horizon and an event horizon. Causally propagated signals inside the event horizon cannot reach future asymptotic observers. In such models one can ask how large can the difference between the area of the event horizon and the area of the apparent horizon be?

Apparent horizons can also appear outside the event horizon in non-Einstein models of gravity. Stationary solutions are the same in Einstein and Brans-Dicke theory \cite{Hawking:1972tf}. If the spacetime is globally static then the event horizon and apparent horizon will coincide. However, in dynamical situations the apparent horizon can appear outside the event horizon and the event horizon area can decrease in the Jordan-string frame in Brans-Dicke theory \cite{Scheel:1994yn}. This is because the null energy condition can be violated in this frame even though it is not violated in the Einstein frame. The location of the apparent horizon changes under a conformal transformation of the metric. The location of the event horizon does not change.

The change in area of the event horizon is related to the expansion. The change of the expansion of the event horizon responds to the local geometry through the Raychaudhuri equation and $R_{ab}l^{a}l^{b}$. But the location of the event horizon is wholly non-local as is the requirement that it lie outside the apparent horizon. But the area is decreasing when $\theta_{l} < 0$ and this is also a condition for it to be inside the apparent horizon. The null energy condition needs to be violated locally to drive the expansion of the event horizon down from a positive value to a negative value to allow it to pass inside the apparent horizon.

The area of the trapping horizon can also be increasing even when the local geometry is static vacuum. This is because the horizon can acquire shear. The shear obtains from the requirement that the surface be closed. Imagine the case where a thin, high speed, pencil of matter is sent into a black hole. The trapping horizon will increase in area even in regions that are initially causally disconnected from the incoming matter, since the horizon develops shear and shear is enough to increase the area of the trapping horizon. Shear is a property of the null rays normal to the surface and hence depends on the surface, not the local geometry.

To study these issues we will present a number of different cases. Firstly, we will examine the case of black holes that are accreting or evaporating at a constant rate. With the implicit assumption that an event horizon does exist in the future, the constant rate case allows certain conditions to be applied that enable us to solve for the location of the event horizon without needing to solve for the full spacetime. In these simple cases we can derive analytical formulae relating the areas of the event horizon and trapping horizons.

To extend this analysis to more realistic situations we will consider the case of a black hole that transitions from growing through matter accretion to evaporation. This simple model is motivated by considering an isolated astrophysical black hole that is initially accreting energy from the cosmic microwave background (CMB) until the temperature of the CMB drops below that of the black hole and it starts to lose mass through Hawking radiation. In this situation we will solve for the location of the event horizon numerically assuming that the situation settles down in the far future to an almost constant rate evaporating black hole. During the transition from accretion to evaporation we will be able to follow the evolution of the apparent horizon and event horizon and see that the apparent horizon moves outside the event horizon before the black hole finishes accreting.

In the third part of the paper we will consider the issue of conformal transformations. For a specific example we will consider black holes in Brans-Dicke theory. We will show how the definition of a trapping horizon can be modified to enable its location to be invariant under a conformal transformation. Interestingly, this condition can be related to black hole thermodynamics and the gravitational entropy of the black hole horizon. We also show how this new definition guarantees the entropy increase theorem in Brans-Dicke theory.

\section{Location of horizons}

The trapping horizons are three-dimensional surfaces, foliated by closed spacelike two surfaces for which the future directed null normals $l^{a}$ and $n^{a}$ satisfy
\bea \theta_{l} & = & 0 \nonumber \\
\theta_{n} & < & 0 \nonumber \\
{\cal{L}}_{n}\theta_{l} & < & 0 \eea
There is an implicit condition here in the choice of $l$ and $n$. Each choice of $l$ and $n$ defines a set of spacelike surfaces normal to them. Different choices of $l$ and $n$ will lead to trapping horizons at different locations. This issue has been examined in \cite{Nielsen2010} and can be formulated in terms of a choice of spacetime slicing into spacelike hypersurfaces. A general spherically symmetric metric can be written in advanced null Eddington-Finkelstein coordinates ($v,r,\theta,\phi$) as
\beq \label{sphersymmetric} \d s^{2} = -e^{-2\Phi(v,r)}\triangle(v,r)\d v^{2} + 2e^{-\Phi(v,r)}\d v\d r + r^{2}\d\Omega^{2} \eeq
In this general spherically symmetric spacetime, for spherically symmetric slicings the trapping horizon null normals will be radial null vectors and the location of the trapping horizon is just given by
\beq \triangle(v,r) = 0 \eeq
This is a spacelike surface for $\dot{m} > 0$, a null surface for $\dot{m}=0$ and a timelike surface for $\dot{m} < 0$. Choosing the conventional parametrization in terms of the Misner-Sharp mass, $\triangle(v,r) = 1-2m(v,r)/r$, the condition for the horizon is just
\beq r=2m(v,r) \label{r_is_two_m} .\eeq
This implicit equation can be solved explicitly in simple cases such as the Vaidya spacetime where $m(v,r)=m(v)$. For a linear mass function of the form $m(v) = m_{o} + \dot{m}v$, in terms of the timelike coordinate $t=v-r$, the horizons will be located at
\beq r = \frac{2m_{o} + 2\dot{m}t}{1-2\dot{m}} \label{r_is_not_two_m}. \eeq
Due to results in \cite{Ashtekar:2005ez} we expect that the non-spherically symmetric trapping horizons will intersect the spherically symmetric ones. For further details see \cite{Nielsen2010}. Since it is spherically symmetric the trapping horizon located at $r=2m(v,r)$ has area $4\pi r^{2} = 16\pi m(v,r)^2$. 

The event horizon is defined as a connected components of the past causal boundary of future null infinity and is generated by null geodesics that fail to reach infinity. The event horizon is always a null surface since it is a causal boundary. In the above spacetime (\ref{sphersymmetric}) the coordinate $v$ is constant on ingoing radial null geodesics. Any outgoing radial null geodesic must satisfy
\beq \frac{\d r}{\d \tau} = \frac{e^{-\Phi}\triangle}{2}\frac{\d v}{\d \tau} \eeq
for some parameter $\tau$ along the curve. In particular, the null generators of the event horizon must satisfy this condition. This first order ordinary differential equation generates the path of all outgoing radial null geodesics. In order to give the location of the event horizon it requires a boundary condition that corresponds to the known location of the event horizon at some particular point. In practice this is usually given by the position of the event horizon at some future point, either when the black hole evaporates entirely or settles down to a stationary state. If the black hole at some point settles down to a Schwarzschild black hole with no further matter accreting, then the event horizon can be located by tracing back the null rays from the future Schwarzschild radius.

\section{Constant rate case}

The event horizon likewise has area $4\pi r_{_{EH}}^{2}$ on spherically symmetric slicings. The difference in the areas $A_{_{diff}}$ is therefore
\beq A_{_{diff}} = 4\pi r_{_{EH}}^{2} - 16\pi m(v,r)^2 \eeq
For the case where the black hole is evolving (either growing or shrinking) at a constant rate we may guess that the difference in the areas of the trapping horizon and event horizon should be constant. With respect to a slicing with slices labelled by $\tau$
\beq \label{constantarea} \frac{\d}{\d \tau}A_{_{diff}} = 0 \eeq
This gives
\beq 2r_{_{EH}}\frac{\d r_{_{EH}}}{\d \tau} - 8 m_{_{AH}}\frac{\d m_{_{AH}}}{\d \tau} = 0 \eeq
substituting in for $\d r_{_{EH}}/\d \tau$
\beq \label{ehrad} r_{_{EH}} = 2m_{_{EH}} + 4r_{_{AH}}\dot{m}_{_{AH}}e^{\Phi_{_{EH}}} \eeq
where $\dot{m} = \frac{\d m}{\d v} = \frac{\d m}{\d \tau}\frac{\d \tau}{\d v}$. Since we are comparing areas on a slice of constant $\tau$, we can expand values at the event horizon in terms of values at the apparent horizon and $r$
\beq m_{_{EH}} = m_{_{AH}} + m'_{_{AH}}(r_{_{EH}}-r_{_{AH}}) + \frac{m''_{_{AH}}}{2}\left(\delta r\right)^{2} + ... \eeq
\beq \Phi_{_{EH}} = \Phi_{_{AH}} + \Phi'_{_{AH}}(r_{_{EH}}-r_{_{AH}}) + \frac{\Phi''_{_{AH}}}{2}\left(\delta r\right)^{2} + ... \eeq
where $\delta r = (r_{_{EH}}-r_{_{AH}})$. Substitute these expansions in to (\ref{ehrad}) to get
\beq r_{_{EH}} = r_{_{AH}} + \frac{m''_{_{AH}}}{\left(1-2m'_{_{AH}}\right)}\left(\delta r\right)^{2} + ... + \frac{4r_{_{AH}}\dot{m}_{_{AH}}e^{\Phi_{AH}}}{\left(1-2m'_{_{AH}}\right)}\left(1+\Phi_{_{AH}}'\delta r + ...\right) \eeq
which is just
\beq \delta r = \frac{m''_{_{AH}}}{\left(1-2m'_{_{AH}}\right)}\left(\delta r\right)^{2} + \frac{2\dot{m}_{_{AH}}}{\kappa_{_{AH}}}\left(1+\Phi_{_{AH}}'\delta r + \frac{1}{2}\left( \Phi''_{_{AH}} + \Phi'^{2}_{_{AH}}\right)(\delta r)^{2}\right) + {\cal{O}}\left( (\delta r)^{3}\right) \eeq
where $\kappa$ can be interpreted as a dynamical surface gravity \cite{Nielsen:2007ac}. Now solve this at linear order in $\delta r$
\beq \delta r = \frac{2\dot{m}}{\kappa}\left(1 + \frac{2\dot{m}}{\kappa}\Phi' \right) + {\cal{O}}(\dot{m}^3) \eeq
and thus, for $\kappa > 0$ to leading order in $\dot{m}$ we expect that the event horizon will be outside the spherically symmetric trapping horizon for $\dot{m} > 0$ but inside for $\dot{m} < 0$.

In the case where the black hole is accreting matter at a steady rate and is a suitably long way from changing to a different state one can also find the approximate location of the event horizon by imposing the condition
\beq \label{ehlocation} \frac{\d^{2}r}{\d v^{2}} = 0 .\eeq
on the horizon generators. This just reflects the fact that the event horizon is growing at a steady rate. While the condition (\ref{constantarea}) compares values at two different points in the spacetime, this condition is purely local on the event horizon. In this case, equation (\ref{ehlocation}) has the general solution
\beq \label{steadyeh} r = \frac{m(v)}{4\dot{m}}\left(1-\sqrt{1-16\dot{m}}\right) .\eeq
which holds at the event horizon. For $\dot{m} \ll 1$ this gives
\beq r = 2m(v)\left(1 + 4\dot{m} + 32\dot{m}^2 + {\mathcal{O}}(\dot{m}^3)\right) \eeq
In terms of the coordinate $t=v-r$ for the linear mass function, equation (\ref{steadyeh}) can be solved explicitly and the event horizon has radial coordinate
\beq r \sim \frac{2m_{o}+2\dot{m}t}{1-2\dot{m}} + \frac{8m_{o}\dot{m}}{1-2\dot{m}}.\eeq
This is just the location of the spherically symmetric trapping horizon with a constant offset of $8m_{o}\dot{m}$ provided $\dot{m} \ll 1$. In this approximation the generators of both the trapping horizon and the event horizon have the same components but the norm of the generators is $4\dot{m}$ for the trapping horizon and zero for the event horizon. The trapping horizon is spacelike for $\dot{m} > 0$ but the event horizon is still a null hypersurface.

\section{Differences in areas}

In terms of the difference between the areal radius coordinates, $\delta r$, defined in the last section, the difference in the areas is
\beq A_{diff} = 4\pi\left(2r_{AH}\delta r - \delta r^{2}\right) \eeq 
To get an idea of the numbers involved in the dynamical accretion and evaporation of mass by a black hole consider the following very approximate situations. The Eddington limit gives a good approximation  of the rate at which black hole accretion disks can be supported by their own self-generated radiation pressure. The Eddington limit is given by
\beq \frac{\d E}{\d t} \simeq 1.3 \times 10^{31}\left(\frac{M}{M_{\odot}} \right) \mathrm{W} \eeq
The dimensionless mass accretion rate is calculated by
\beq \dot{m} = \frac{G}{c^{5}}\frac{\d E}{\d t} \eeq
For typically observed values of a black hole accreting at a tenth of the Eddington rate \cite{Kollmeier:2005cw} the dimensionless accretion rate is approximately
\beq \dot{m} \simeq 10^{-22}\left( \frac{M}{M_{\odot}} \right) .\eeq
The matter falling into the black hole is ten times the energy being emitted as light. Since this accretion is usually associated with a disk it will not be exactly spherically symmetric, but the approximation is still often applied.

For a black hole accreting purely from the Cosmic Microwave Background which is assumed to be isotropic, and using the Stefan-Boltzmann law, we have approximately that the area, $A$, is
\beq A \simeq 16\pi\frac{G^{2}}{c^{4}}M_{\odot}^{2}\left( \frac{M}{M_{\odot}} \right)^{2} \eeq
and so the dimensionless mass accretion rate is
\beq \dot{m} \simeq 10^{-50}\left(\frac{T}{T_{3K}}\right)^{4}\left(\frac{M}{M_{\odot}}\right)^{2} .\eeq
For a black hole whose dynamics are dominated by evaporation through Hawking radiation with temperature
\beq T_{_{\mathrm{BH}}} = \frac{1}{8\pi}\frac{c^{3}\hbar}{k_{_{\mathrm{B}}}GM} \eeq
we have a (negative) massless accretion rate of
\beq \label{mdotHawk} \dot{m} \simeq -10^{-81}\left(\frac{M_{\odot}}{M}\right)^{2} .\eeq
The difference in areas can be computed as
\beq \label{Adiff} A_{_{diff}} = A_{_{EH}} - A_{_{AH}} \simeq 10^{78}\dot{m}\left(\frac{M}{M_{\odot}}\right)^{2}l_{_{P}}^{2} \eeq
The numerical factor is related to the solar Schwarzschild area in Planck units, $4.2\times 10^{77}$. The exact formula for the area difference for the case where the black hole is evaporating purely through Hawking radiation ((\ref{mdotHawk}) into (\ref{Adiff})) turns out to be
\beq A_{_{diff}} = -\frac{1}{120}\frac{G\hbar}{c^{3}} = -\frac{1}{120}l_{_{P}}^{2} \eeq
Thus in the case where an otherwise isolated black hole is evaporating at an almost constant rate due to Hawking radiation, in the conventional picture of black hole evaporation \cite{Hawking:1974sw}, the difference in the area of the event horizon and trapping horizon is not resolvable at the Planck scale (although the trapping horizon is outside the event horizon). For any of the accreting black holes considered above, the difference is many orders of magnitude of the Planck area.

For a solar-mass black hole accreting at a tenth of the Eddington rate the difference in areas between the event horizon and the spherically symmetric trapping horizon will be around $10^{56}$ in units of Planck area, while for a supermassive black hole of mass $10^{8}$ solar masses, accreting purely form the CMB, the difference in areas will be around $10^{60}$ in Planck units.

\section{Varying rate case}

To see the transition of a trapping horizon from inside the event horizon to outside, we need to look at black holes where the rate of change is changing. One example of this is an otherwise isolated black hole that accretes matter from the cosmic microwave background that fills all of the universe. A static solar-mass sized black hole has a Hawking temperature of about $T_{BH} = 10^{-8}$K. The temperature of the CMB is currently about $T_{CMB}=3$K. The temperature of the black hole changes as its mass changes and the temperature of the CMB changes as the universe expands. Roughly speaking, if the temperature of the CMB is larger than the temperature of the black hole, it will accumulate mass and grow. If the temperature of the black hole is greater than the temperature of the CMB it will lose mass and shrink.

Black holes of mass $<10^{23}$kg have a temperature today greater than the CMB. Primordial black holes of initial mass $\sim 10^{13}$kg should be evaporating today \cite{Carr:2005zd}. Their mass density however must be $\Omega < 10^{-8}$ \cite{Carr:2009jm}.

If we assume that the universe is dominated by a cosmological constant with $\Lambda \sim 10^{-35}\mathrm{s}^{-2}$ the temperature of the CMB will equal $10^{-8}$K in about $10^{18}$seconds or $10^{11}$years. During this time the black hole will accrete about $1 ~$kg of photonic matter from the CMB. Since the mass of a solar-sized black hole is about $10^{30}$kg it's percentage increase in mass will be tiny.

In a similar length of time the mass lost to Hawking radiation will be even smaller, about $10^{-25}$kg. Therefore, during the time that the CMB temperature falls to the black hole temperature, one can consider the black hole size as roughly constant.


Consider a simple toy model consisting of a spherically symmetric black hole that absorbs radiation from the CMB and also emits Hawking radiation. One way of modelling the Hawking radiation is to implement it as inflowing negative energy \cite{Hiscock:1980ze} or with a region of outflowing positive energy but negative energy onto the black hole \cite{Hayward:2005gi}. The Vaidya solution for infalling radiation is
\beq \d s^{2} = -\left(1-\frac{2m(v)}{r}\right)\d v^{2} + 2\d v\d r + r^{2}\d\Omega^{2} \eeq
For a black hole that accretes purely from the CMB and then evaporates via Hawking radiation we can write a mass function as
\beq m(v) = m_{o} - k_{1}(v-v_{o}) - k_{2}\mathrm{exp}\left(\frac{-(v-v_{o})}{\lambda}\right) + k_{2} \eeq
where $k_{1}$, $k_{2}$ and $\lambda$ are constants. The second term given the mass lost due to Hawking radiation and the third term gives the mass gained by accretion from the CMB in a de Sitter (cosmological constant dominated) universe. Interpreting the terms as accretion from the CMB and evaporation through Hawking radiation we would have
\beq k_{1} = \frac{\sigma}{256}\frac{1}{m_{o}^2} \eeq
\beq k_{2} = 4\pi\sigma m_{o}^{2}T_{o}^{4}\sqrt{\frac{3}{\Lambda}} \eeq 
\beq \lambda = 4\sqrt{\frac{\Lambda}{3}} \eeq
where $\sigma$ is the Stefan-Boltzmann constant, $m_{o}$ is the mass of the black hole at $v=v_{o}$, $T_{o}$ is the temperature of the CMB at $v=v_{o}$ and $\Lambda$ is the cosmological constant. Since it is based on the Vaidya solution this model does not embed the black hole in a true cosmological background either de Sitter-like or otherwise and is only intended to hold in a small region near the horizons to illustrate the potential behaviour in more exact models. The accretion from CMB part acts like a heat bath at a given temperature rather than true accretion from a Robertson-Walker spacetime.

The trapping horizon reaches its maximum area where $\dot{m}=0$. This corresponds to
\beq v_{_{\mathrm{THmax}}}-v_{o} = \lambda\ln\left(\frac{k_{2}}{\lambda k_{1}}\right) \eeq
The trapping horizon crosses the event horizon roughly when
\beq v_{_{\mathrm{cross}}} - v_{o} = \lambda \ln \left(\frac{k_{2}}{4m_{o}k_{1}}\right) \eeq
Therefore the time (elapse of the $v$ coordinate) between the trapping horizon passing outside the event horizon and the area of the trapping horizon starting to decrease (where the NEC is locally violated), is
\beq v_{_{\mathrm{THmax}}} - v_{_{\mathrm{cross}}} = \lambda\ln\left(\frac{4m_{o}}{\lambda}\right) \eeq
\begin{figure}[!h]
\centering
\includegraphics[width=0.98\columnwidth,clip]{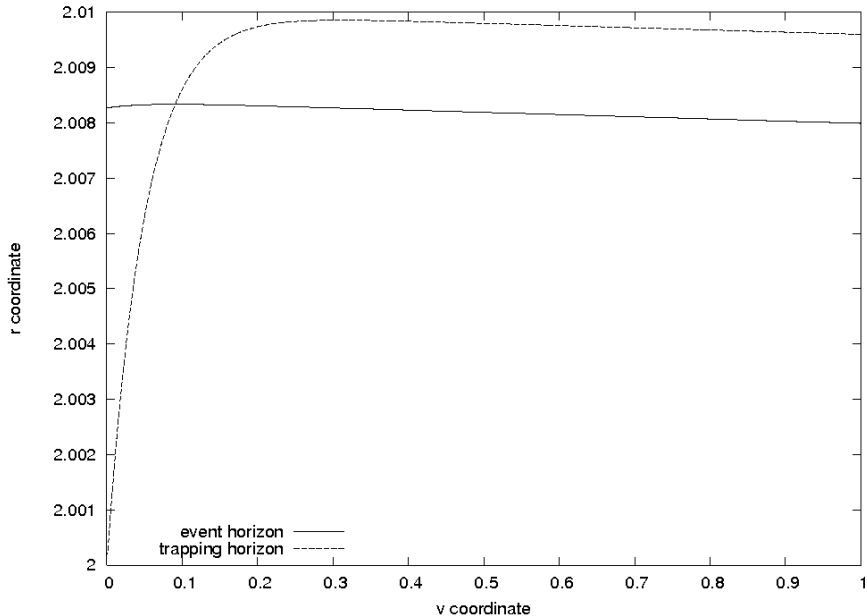}
\caption{The coordinates of the event horizon (solid line) and trapping horizon (dashed line) for an example model with $m_{o} = 1.0$ $k_{1}=0.0002$, $k_{2}=0.005$ and $\lambda = 0.05$. The trapping horizon moves outside the event horizon while the trapping horizon is still growing.}
\label{fig:cosmohorizon}
\end{figure}
An example for the case $m_{o} = 1.0$ $k_{1}=0.0002$, $k_{2}=0.005$ and $\lambda = 0.05$ is shown in Fig.(\ref{fig:cosmohorizon}). The location of the event horizon is found by tracing back from the value given by (\ref{steadyeh}). The difference in radial coordinate at $v\sim 1$ is due to the fact that the black hole is still dynamical when it settles down to ``constant'' evaporation.

In this example the trapping horizon starts off inside the event horizon and grows rapidly as the black hole accretes matter. The trapping horizon crosses the event horizon just as the event horizon starts to decrease in area. In the example this occurs at $v=0.09$ and $m(v)=1.0042$. At this point the accreting ``CMB'' flux is still higher than the ``evaporation'' flux and we have $\dot{m} > 0$ implying that the null energy condition is still satisfied. The apparent horizon grows to a maximum size of $r=2.0099$ before the ``evaporation'' flux starts to dominate and the hole starts to shrink.

This example shows that it is possible, even likely, that the trapping horizon will appear outside the event horizon before the trapping horizon starts to shrink. The crossing occurs precisely where the area of the event horizon starts to decrease. This means that the trapping horizon becomes visible to asymptotic observers even before the energy conditions are violated and before its dynamics are dominated by Hawking radiation. If the trapping horizon starts to produce Hawking radiation in its near vicinity after this time the Hawking radiation can escape to infinity. The event horizon's area starts to decrease, not because of some causal signal such as local violation of the energy conditions, but in anticipation of future violation. This is then consistent with the violation of energy conditions being associated with a Hawking flux from the trapping horizon and mass loss to infinity which causes the black hole to shrink.

\section{Rescaling}

Under a conformal transformation the relative position of the event horizon and the trapping horizon can also change. The use of conformal transformations is fairly common in looking at black holes solutions. This is particularly true in string theory where conformal transformations are used to relate the string frame, with a non-minimally coupled dilaton field, to the Einstein frame. It has been argued in the literature that classically the two frames are physically equivalent \cite{Faraoni:2006fx}. The conformal transformation has the effect of rescaling lengths and areas as measured by the metric. The physical effect of this rescaling is, for example, to change the meaning of mass since the norm of the four-momentum, $p^{a}p_{a}$, will no longer be constant from point to point or time to time. 

It has been observed in numerical simulations of black hole collapse in Brans-Dicke theory, that the trapping horizon can appear outside the event horizon \cite{Scheel:1994yn}. This occurs despite the fact that the string frame of Brans-Dicke theory can be related via a conformal transformation to Einstein theory with a scalar field that obeys the null energy condition. In the Einstein frame the trapping horizon appears exclusively behind the event horizon, in accordance with the theorem of Hawking and Ellis \cite{Hawking:1973uf}.

Two issues are involved here. Firstly, unlike the event horizon, the location of the trapping horizon changes under a conformal transformation \cite{Scheel:1994yn}. Secondly, the trapping horizon can appear outside the event horizon in the string frame because the Einstein equations do not hold in this frame.

To relate the two frames the metric is scaled by a conformal factor that can vary with spacetime point
\beq \label{conformaltrans} g_{ab} \rightarrow g'_{ab} = W(x)g_{ab} \eeq
The expansion of a null vector $l^{a}$ in any frame is given by
\beq \theta_{l} = q^{ab}\nabla_{a}l_{b} = \left( g^{ab} + \frac{l^{a}n^{b}}{\left( -n^{c}l^{d}g_{cd}\right)} + \frac{n^{a}l^{b}}{\left( -n^{c}l^{d}g_{cd}\right)}\right)\nabla_{a}l_{b} \eeq  
where $q^{ab}$ is a projection tensor onto the two-dimensional spacelike surface to which $l^{a}$ and $n^{a}$ are normal. (If $l^{a}$ is defined as globally null then the third term vanishes identically.) This is a result that holds for a Lorentzian signature manifold independently of whether the Einstein equations hold. In general there is a freedom to rescale null vectors even without rescaling the metric. The vanishing of the expansion does not depend on a pure rescaling of the null vector $l^{a} \rightarrow Wl^{a}$, although its value does since under this rescaling we have
\beq \theta_{l} \rightarrow W\theta_{l} \eeq
Under a conformal transformation of the form (\ref{conformaltrans}) we have ${\tilde{g}}^{ab}=W^{-1}g^{ab}$ and $q^{ab} \rightarrow W^{-1}q^{ab}$. We can fix the normalization of $l^{a}$ by requiring ${\tilde{l}}^{a} = l^{a}$ with ${\tilde{l}}_{a} = Wl_{a}$ and thus
\beq {\tilde{\nabla}}_{a}{\tilde{l}}_{b} = W\nabla_{a}l_{b} + l_{b}\nabla_{a}W - \frac{1}{2}\left(l_{a}\nabla_{b}W + l_{b}\nabla_{a}W - g_{ab}l^{c}\nabla_{c}W\right) \eeq
therefore
\beq \label{thetaltrans} {\tilde{\theta}}_{l} = \theta_{l} + \frac{l^{a}\nabla_{a}W}{W} \eeq
The vanishing of $\theta_{l}$ for a given surface is therefore not necessarily invariant under a conformal transformation. The conformal transformation though does not change the coordinates of a given spacetime event nor the path of null rays. The location of the event horizon is unchanged. In one frame the solution of $\theta_{l}=0$ may lie inside the event horizon and in another frame outside.

Brans-Dicke theory is the prototype alternative to theory of gravity with scalar and tensor modes. The action, in the string frame is given by
\beq {\cal{L}} = \frac{1}{16\pi}\left( \phi R - \frac{\omega}{\phi}\nabla_{a}\phi\nabla^{a}\phi\right) + {\cal{L}}_{matter} .\eeq
Variation of this action with respect to the metric gives the gravitational field equations
\beq G_{ab}\phi = 8\pi T^{matter}_{ab} + \frac{\omega}{\phi}\left(\nabla_{a}\phi\nabla_{b}\phi - \frac{1}{2}g_{ab}\nabla_{c}\phi\nabla^{c}\phi \right) + \nabla_{a}\nabla_{b}\phi - g_{ab}\nabla_{c}\nabla^{c}\phi \eeq
The proof of the apparent area theorem \cite{Hawking:1973uf} is a purely geometric proof that only relies on the validity of the Null Raychaudhuri equation and the geometrical condition $R_{ab}l^{a}l^{b} > 0$, the null curvature condition. This condition can be related via the Einstein equations to the null energy condition, $T_{ab}l^{a}l^{b}$. Contracting the Ricci tensor with $l^{a}$ gives
\beq R_{ab}l^{a}l^{b} = \frac{8\pi}{\phi} T_{ab}l^{a}l^{b} + \frac{\omega}{\phi^{2}}\left(l^{a}\nabla_{a}\phi\right)^{2} + \frac{l^{a}l^{b}\nabla_{a}\nabla_{b}\phi}{\phi} \eeq
where $T_{ab}$ is the energy-momentum tensor of the non-gravitational matter fields. Even if the matter obeys the null energy condition $T_{ab}l^{a}l^{b}\geq 0$, the sign of the last term is indeterminate and therefore we may have a violation of the null curvature condition. This is in fact required to happen for the surfaces found in \cite{Scheel:1994yn}. Brans-Dicke theory in the string frame can be recast in the Einstein frame via the conformal transformation
\beq \label{bdconftrans} {\tilde{g}}_{ab} = \phi g_{ab} \eeq
In the Einstein frame the null tangent vectors are unchanged ${\tilde{l}}^{a} = l^{a}$ and they are null with respect to the new metric as well as the old one. In the Einstein frame the gravitational field equations are
\beq {\tilde{G}}_{ab} = 8\pi {\tilde{T}}_{ab} +\frac{3+2\omega}{16\pi \phi^{2}}\left({\tilde{\nabla}}_{a}\phi{\tilde{\nabla}}_{b}\phi - \frac{1}{2}{\tilde{g}}_{ab}{\tilde{\nabla}}_{c}\phi{\tilde{\nabla}}^{c}\phi \right) \eeq
and thus
\beq {\tilde{R}}_{ab}l^{a}l^{b} = 8\pi {\tilde{T}}_{ab}l^{a}l^{b} + \frac{3+2\omega}{16\pi\phi^{2}}l^{a}{\tilde{\nabla}}_{a}\phi l^{b}{\tilde{\nabla}}_{b}\phi \eeq
Provided the matter obeys the null energy condition and $\omega > 0$ the geometry will also obey the null curvature condition in the Einstein frame.

The vanishing of the expansion is equivalent to the statement that the area is unchanged under translations along $l^{a}$ via the relation $l^{a}\nabla_{a}A - \theta_{l}A = 0$. Since the conformal factor changes how areas are measured this no longer selects out the same horizon. The area two-form changes as $q_{ab} \rightarrow Wq_{ab}$. The condition that the Lie derivative of this ``conformally transformed area'' be zero is
\beq {\cal{L}}_{l}\left( W A\right ) = WA\left( \theta_{l} + \frac{{\cal{L}}_{l}W}{W}\right) = 0 \eeq
This is the same as the transformation in (\ref{thetaltrans}). In theories with gravitationally non-minimally coupled scalar fields, such as Brans-Dicke theory, the gravitational entropy, defined by the Noether-charge entropy \cite{Wald:1993nt}, is not always equal to the area. Thermodynamically, it has been argued that in Brans-Dicke theory the quantity $A\phi$ is non-decreasing on the event horizon \cite{Kang:1996rj}. This suggests that in non-minimally coupled theories (or in a frame where the metric tensor doesn't satisfy the Einstein equations) the horizons should be defined not by the vanishing of the area along normal null directions, but by the vanishing of the gravitational entropy, $S_{g}$, along null directions. The three conditions for a trapping horizon would then be
\bea l^{a}\nabla_{a}S_{g} & = & 0 \nonumber \\
n^{a}\nabla_{a}S_{g} & < & 0 \nonumber \\
n^{a}\nabla_{a}\left(l^{a}\nabla_{a}S_{g}\right) & < & 0 \eea
This then provides a frame covariant way to identify the ``true'' trapping horizon. In the Einstein frame this would reduce to the ordinary requirements on the null expansions for a trapping horizon. It has been claimed  that if a field redefinition can be used to relate the actions of two theories then the black hole entropies in these theories are related by the same field redefinition, at least when applied to event horizons \cite{Jacobson:1993vj}. If this is also true of the gravitational entropies of trapping horizons then the above gives a frame independent way of identifying the physical trapping horizon in all theories for which a gravitational entropy can be defined. 

With these conditions one can then examine how the generalised Noether-charge entropy evolves along a horizon with tangent $r^{a} = Bl^{a} + Cn^{a}$. The variation of the generalised entropy is
\beq r^{a}\nabla_{a}S = Bl^{a}\nabla_{a}S + Cn^{a}\nabla_{a}S \eeq
The first term on the right hand side is zero by assumption. Since we require the tangent $r^{a}$ to generate evolution along the generalised trapping horizon we have
\beq r^{a}\nabla_{a}\left( l^{a}\nabla_{a}S\right) = 0 \eeq
This can be rearranged to give
\beq C = -\frac{Bl^{a}\nabla_{a}\left( l^{a}\nabla_{a}S\right) }{n^{a}\nabla_{a}\left( l^{a}\nabla_{a}S\right)} \eeq
and so the change of the generalised entropy along the horizon can be written as
\beq r^{a}\nabla_{a}S = -\frac{Bn^{a}\nabla_{a}S}{n^{a}\nabla_{a}\left(l^{a}\nabla_{a}S\right)}l^{a}\nabla_{a}\left(l^{a}\nabla_{a}S\right) \eeq
With the sign of $B$ setting the orientation of $r^{a}$ and the sign of $n^{a}\nabla_{a}S$ and $n^{a}\nabla_{a}\left(l^{a}\nabla_{a}S\right)$ both negative by assumption on the horizon, whether the generalised entropy is increasing or decreasing is just determined by the last term, $l^{a}\nabla_{a}\left(l^{a}\nabla_{a}S\right)$. In the Brans-Dicke case we have $S=\phi A$. For the entropy to be positive we require $\phi A > 0$ and since the area is positive, $A > 0$, we see that we require $\phi > 0$ too (note that if $\phi = 0$ anywhere in the spacetime then the conformal transformation (\ref{bdconftrans}) becomes illdefined.) The variation of this entropy in the outgoing null direction is then
\beq l^{a}\nabla_{a}S = A\left(\phi\theta_{l} + l^{a}\nabla_{a}\phi\right) \eeq
and
\beq l^{a}\nabla_{a}\left(l^{a}\nabla_{a}S\right) = -A\phi\left(\frac{3}{2}\theta_{l}^{2} + \sigma^{2} + \frac{8\pi G}{\phi}T_{ab}l^{a}l^{b} \right) - \frac{A\omega}{\phi}\left( l^{a}\nabla_{a}\phi \right) ^{2} \eeq
Thus for $\omega > 0$ and matter obeying the null energy condition, $T_{ab}l^{a}l^{b} > 0$, the generalised entropy is guaranteed to increase along the generalised trapping horizon.
\section{Conclusion}

We have seen how the location of the trapping horizon can be related to the location of the event horizon for several different cases. For accreting black holes the difference in areas can be very large relative to the Planck scale. This means that the microscopic degrees of freedom responsible for this entropy are likely to be very different in these cases. For evaporating black holes the difference in area is not resolvable on the Planck scale. 

We have also presented a simple model of a black hole that transitions from accreting to evaporating. The model is inspired by an otherwise isolated black hole accreting from the CMB and subsequently beginning to evaporate as the CMB temperature falls. In this case the trapping horizon appears outside the event horizon before the evaporation becomes dominant and before the null energy condition is violated.

However, both these cases assume the standard picture of black hole evaporation that includes both an event horizon and a trapping horizon. In some models there is only a trapping horizon and no event horizon. Whether there is an event horizon or not is virtually undecidable by experiment though, which is one of the key reasons why some authors have preferred to concentrate on trapping horizons.

We have also suggested a redefinition for trapping horizons outside of the Einstein conformal frame. The definition relies on the notion of dynamical gravitational entropy. In non-Einstein frames the generalised trapping horizon will no longer be a marginally outer trapped surface.

\end{document}